# Quasi-Particle Tunneling in Anti-Pfaffian Quantum Hall State

Toru Ito[*], Kentaro Nomura[1], and Naokazu Shibata

*Department of Physics, Tohoku University, Aoba, Aoba-ku, Sendai 980-8578*
*Institute for Materials Research, Tohoku University, Katahira, Aoba-ku, Sendai[1] 980-8577*

We study tunneling phenomena at the edge of the anti-Pfaffian quantum Hall state at the filling factor $\nu = 5/2$. The edge current in a single point-contact is considered. We focus on nonlinear behavior of two-terminal conductance with the increase in negative split-gate voltage. Expecting the appearance of the intermediate conductance plateau we calculate the value of its conductance by using the renormalization group (RG) analysis. Further, we show that non-perturbative quasi-particle tunneling is effectively described as perturbative electron tunneling by the instanton method. The two-terminals conductance is written as a function of the gate voltage. The obtained results enable us to distinguish the anti-Pfaffian state from the Pfaffian state experimentally.

KEYWORDS: fractional, quantum Hall, Moore-Read, Pfaffian, anti-Pfaffian, $\nu=5/2$, non-abelian, instanton, point contact

The nature of the fractional quantum Hall (FQH) effect at the filling factor $\nu = 5/2$ has attracted much attention because Laughlin and Jain's hierarchy theory does not explain the FQH state at this filling.[1–5] Although several candidates for the ground state at $\nu = 5/2$ have been proposed, many numerical studies have shown that the most successful one is the Moore-Read Pfaffian state.[6–9] The realization of the Pfaffian state is important since quasi-particle excitations from this state obey non-abelian braiding statistics, which could put quantum computing into practice.[10] Although it is considered to be the ground state at the half filling in the absence of Landau level mixing, the Pfaffian state does not have particle-hole symmetry. Recently, the particle-hole conjugate state of the Pfaffian state, which is called the anti-Pfaffian state, has been suggested. This state is also a strong candidate for the ground state.[11,12] Although these two states are degenerate in the thermodynamic limit, many factors such as the Landau level mixing can break the particle-hole symmetry and lift the degeneracy. Since one of these states would be the ground state, experimental probes that can distinguish these two states are matter of concern.

Tunneling experiments in point-contact systems have been used to determine the properties of FQH states.[13,14] In this paper, we consider a single point-contact geometry as is shown in Fig. 1. To investigate tunneling transport, we focus on the nonlinearity of the two-terminal conductance, which is determined by the tunneling property at the point-contact. The two-terminal conductance is defined by $\frac{eI_{SD}}{V_3-V_2}$, where $I_{SD}$ is the source-drain current and $V_2$ and $V_3$ are the voltages in the reservoirs 2 and 3 respectively. In the Landauer-Büttiker formula,[15,16] the two-terminal conductance is written as a function of temperature:

$$G = c\frac{e^2}{h} + \lambda T^{-\alpha}, \qquad (1)$$

where $c$ is determined by the property of the edge current and $\lambda$ is a function of tunneling amplitude. The parameter $\alpha$ is decided by the tunneling property of the electron or quasi-particle.

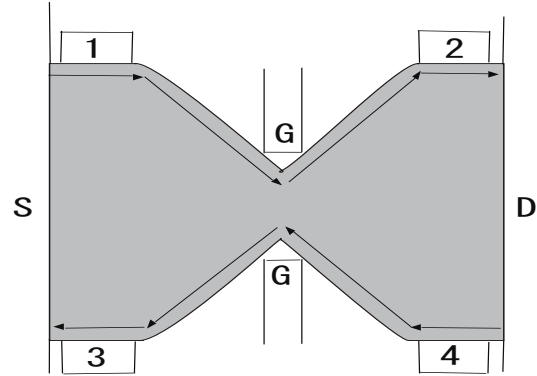

Fig. 1. (Color online)Single point-contact geometry for the FQH system. 1, 2, 3, and 4 are the numbers of voltage probes. S and D are the source and drain terminals. G is the split gate. The two-terminal conductance is defined by $\frac{eI_{SD}}{V_3-V_2}$, where $I_{SD}$ is the source-drain current. The shaded region represents the FQH liquid.

In the following sections, we first introduce the anti-Pfaffian tunneling model. Next, we argue that intermediate conductance plateau is observed by applying a gate voltage, and we calculate the conductance by renormalization group (RG) analysis. Furthermore, by the instanton method we show that non-perturbative quasi-particle tunneling (QPT) is described as perturbative electron tunneling effectively. In summary, the schematic portrait of the conductance of the anti-Pfaffian state is shown as a function of gate voltage.

We consider the model that describes a single point-contact geometry where the Hall-bar between the left and right terminals has upper and lower edges with a scattering point at $x = 0$.[17] We note that we drop the edges of the lowest Landau level of both spins because they have no role in the following discussion. The anti-Pfaffian state is the particle-hole conjugate of the Pfaffian state, thus to describe the anti-Pfaffian edge theory, we start from writing the effective edge

[*]E-mail: i-toru@cmpt.phys.tohoku.ac.jp

Lagrangian for the Pfaffian state.[18,19] It is written as

$$L_{Pf}(\psi,\phi) = \frac{2}{4\pi}\partial_x\phi(i\partial_t + v_c\partial_x)\phi + i\psi(\partial_t + v_n\partial_x)\psi, \quad (2)$$

where the field $\phi$ describes chiral bosonic charge modes and $\psi$ describes the chiral neutral Majorana mode.[19] Here, the charge velocity $v_c$ and the neutral velocity $v_n$ have the same sign, since these modes propagate in the same direction. The edge theory of the anti-Pfaffian state is given by considering the particle-hole conjugate of the Pfaffian state. The anti-Pfaffian wavefunction is defined by $\Psi_{an-Pf} = \hat{Q}_{Pf}\Psi_{\nu=1}$, where $\Psi_{\nu=1}$ is the $\nu = 1$ QH wavefunction and $\hat{Q}_{Pf}$ is the creation operator of the Pfaffian state of holes in the filled Landau level. This equation indicates that under a particle-hole transformation, the boundary between the anti-Pfaffian state and the vacuum ($\nu = 0$) is mapped onto the boundary between the Pfaffian state and the vacuum of a hole ($\nu = 1$). Therefore, the edge at the boundary of the anti-Pfaffian state is described by the Pfaffian edge mode and another chiral boson mode that corresponds to the $\nu = 1$ edge. In this case, there are two channels in one boundary; we therefore should consider the Coulomb repulsion between the two channels. The effective Lagrangian is then written as

$$L = L_{Pf}(\psi_1,\phi_2) + \frac{1}{4\pi}\partial_x\phi_1(-i\partial_t + v_1\partial_x)\phi_1 + \frac{2}{4\pi}v_{12}\rho_1\rho_2. \quad (3)$$

$L_{Pf}$ is given by eq. (2), and $\psi_1$ and $\phi_2$ are the Majorana and boson modes of the Pfaffian edge of a hole respectively. The second term describes the $\nu = 1$ edge, where $v_1$ is the charge velocity and $\phi_1$ is the chiral boson mode of this channel. This mode propagates in the inverse direction of the Pfaffian mode. The last term $v_{12}\rho_1\rho_2$ represents the short-ranged Coulomb repulsion between the two edges, where $\rho_1 = \partial_x\phi_1$ and $\rho_2 = \partial_x\phi_2$ are charge densities. We assume a sufficiently clean sample so that the impurity effect is negligible.[20] To rewrite the Lagrangian into a more simple form, we introduce charge and neutral decompositions, $\phi_\rho = \phi_1 - \phi_2$ and $\phi_\sigma = \phi_1 - 2\phi_2$. The effective Lagrangian of the anti-Pfaffian edge is written as

$$L = \frac{2}{4\pi}\partial_x\phi_\rho(-i\partial_t + v_\rho\partial_x)\phi_\rho + i\psi_1(\partial_t + v_n\partial_x)\psi_1$$
$$+ \frac{1}{4\pi}\partial_x\phi_\sigma(i\partial_t + v_\sigma\partial_x)\phi_\sigma - \frac{2}{4\pi}v_{\rho\sigma}\partial_x\phi_\rho\partial_x\phi_\sigma, \quad (4)$$

where $v_\rho = \frac{v_1+v_c+4v_{12}}{2}$, $v_\sigma = v_1 + v_c + 2v_{12}$, and $v_{\rho\sigma} = v_1 + v_c - \frac{3v_{12}}{2}$. For small $v_{\rho\sigma}$ values, the last term is perturbatively irrelevant[12] and then the Lagrangian is rewritten as

$$L \approx \frac{2}{4\pi}\partial_x\phi_\rho(-i\partial_t + v_\rho\partial_x)\phi_\rho + i\sum_{a=1,2,3}\psi_a(\partial_t + v'_a\partial_x)\psi_a, \quad (5)$$

where we have used fermionizing, $\exp(i\phi_\sigma) = \psi_2 + i\psi_3$. Here, $v'_a$ are $v_n$ for $a = 1$ and $v_\sigma$ for $a = 2, 3$. This form is almost the same as the Lagrangian of the Pfaffian edge, thus we can treat the anti-Pfaffian edge similarly to the Pfaffian edge. We next consider tunneling effect of the electron and quasi-particle. The tunneling Hamiltonian is written by using electron and quasi-particle operators. The electron operator of the anti-Pfaffian edge is determined by the edge Lagrangian.[11,12,18] It is written as

$$\psi_{el} = \sum_{a=1,2,3}\psi_a e^{2i\phi_\rho}. \quad (6)$$

The quasi-particle operator, which is given by the conformal field theory (CFT) is written as[11,12]

$$\psi_{qp} = \psi_{\frac{1}{2}}e^{\frac{i\phi_\rho}{2}}, \quad (7)$$

where $\psi_{\frac{1}{2}}$ is the spin-1/2 field of $SU(2)_2$, and this operator is composed of terms of Ising order and the disorder fields $\sigma_a$ and $\mu_a$ in the CFT of the central charge $c = 1/2$. We note that the sum of $a$ is omitted in the following equations. The tunneling Hamiltonians are written as

$$H_{ET} = \frac{\Gamma_{ET}}{2}(\psi_{el}^{R\dagger}\psi_{el}^L + \psi_{el}^{L\dagger}\psi_{el}^R)$$
$$= \Gamma_{ET}\sum_a\psi_a\bar{\psi}_a\cos(2(\phi_\rho - \bar{\phi}_\rho)), \quad (8)$$

$$H_{QPT} = \frac{\Gamma_{QPT}}{2}(\psi_{qp}^{R\dagger}\psi_{qp}^L + \psi_{qp}^{L\dagger}\psi_{qp}^R)$$
$$= \Gamma_{QPT}\psi_{\frac{1}{2}}\cos(\frac{\phi_\rho - \bar{\phi}_\rho}{2}), \quad (9)$$

where the indexes $L$ and $R$ indicate the left and right moving mode respectively. The parameters $\Gamma_{ET}$ and $\Gamma_{QPT}$ are the tunneling amplitude, and $\psi_{\frac{1}{2}} = \psi_{\frac{1}{2}}^R\psi_{\frac{1}{2}}^L$. These Hamiltonians are composed of CFT operators and have the conformal invariance. Thus, we can easily derive the renormalization group (RG) equation by counting the conformal dimension.[21] The RG equation of quasi-particle tunneling (QPT) is written as

$$\frac{d\Gamma_{QPT}}{dl} = [1 - g]\Gamma_{QPT}, \quad (10)$$

where $g$ is the sum of conformal dimensions $g = \frac{1}{8} + \frac{3}{8}$. $e^l = \frac{\Lambda_0}{\acute{\Lambda}}$, where $\Lambda_0$ and $\acute{\Lambda}$ are bare and renormalized cut-off respectively. In our RG analysis, $k_BT$ plays a role of cut-off. The $\beta$-function is positive, thus we have to study QPT non-perturbatively. In the same way we can calculate the $\beta$-function for $\Gamma_{ET}$, which is negative. Thus, we can treat electron tunneling perturbatively. In the next section, we argue the edge configurations for several situations.

We first consider the small and large limits of the gate voltage in a point-contact.[13] In the small gate voltage limit, configurations of edge channels are shown in Fig. 2(a). Quasi-particle tunneling occurs between upper and lower edges, so this limit is called the weak backscattering limit. In the large gate voltage limit shown in Fig. 2(b), the edge channels are pinched-off and separated by the vacuum. This limit is called the weak tunneling limit since electron tunneling occurs between the left and right FQH liquids. In cases that the edge of FQH is composed of two types of edges, such as $\nu = 2/3$ and $2/5$ edges, it is known that an intermediate plateau, which is different from $\nu e^2/h$, appears between the small and large voltage limits.[22–24] One of edges is in the backscattering limit and the other is in the weak tunneling limit in the intermediate state. Thus only one edge channel contributes to the conductance. In the $\nu = 2/5$ FQH state, for example, $e^2/3h$ conductance plateau is observed with the increase in gate voltage. Here, we study the problem whether an intermediate plateau appears or not in the anti-Pfaffian edge transport regime basing on the RG analysis. To determine which edge contributes to the intermediate conductance, we consider the anti-Pfaffian edge eq. (3) and investigate the edge tunneling of the Pfaffian and $\nu = 1$ QH states. The electron operator of the Pfaffian edge is $\psi_{el-pf} = \psi \exp(-2i\phi)$ and its quasi-particle operator is $\psi_{qp-pf} = \sigma \exp(-i\phi/2)$, thus the Pfaffian edge tunneling Hamiltonians for electron and quasi-particle are written as

$$H_{ET-pf} = \frac{\Gamma_{ET-pf}}{2}(\psi^R_{el-pf}{}^\dagger \psi^L_{el-pf} + \psi^L_{el-pf}{}^\dagger \psi^R_{el-pf})$$
$$= \Gamma_{ET-pf}\psi\bar\psi \cos(2(\phi - \bar\phi)), \qquad (11)$$

$$H_{QPT-pf} = \frac{\Gamma_{QPT-pf}}{2}(\psi^R_{qp-pf}{}^\dagger \psi^L_{qp-pf} + \psi^L_{qp-pf}{}^\dagger \psi^R_{qp-pf})$$
$$= \Gamma_{QPT-pf}\sigma \cos(\frac{\phi - \bar\phi}{2}). \qquad (12)$$

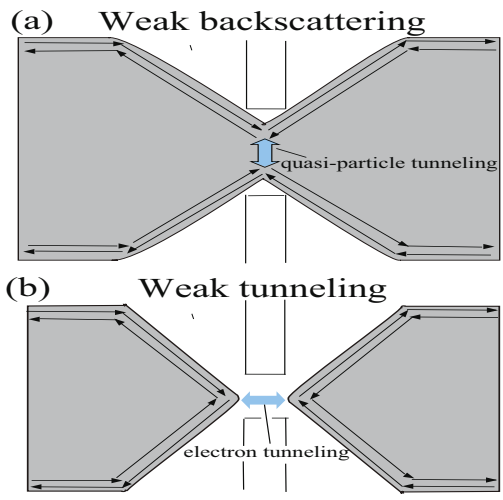

Fig. 2. (Color online) (a)the weak backscattering limit is represented. Quasi-particle tunneling occurs. Here, two edge channels contribute to the constant part of the two-terminal conductance. (b)the weak tunneling limit is represented. Electron tunneling occurs. The edge channels are separated by the vacuum, and thus they do not contribute to the constant part of the conductance. The shaded regions represent the FQH liquid.

By counting the conformal dimension, we obtain the renormalization group equation for the tunneling parameters $\Gamma_{ET-pf}$ and $\Gamma_{QPT-pf}$,

$$\frac{d\Gamma_{ET-pf}}{dl} = [1 - (1+3)]\Gamma_{ET-pf}, \qquad (13)$$

$$\frac{d\Gamma_{QPT-pf}}{dl} = [1 - (\frac{1}{8} + \frac{1}{8})]\Gamma_{QPT-pf}. \qquad (14)$$

For $\nu = 1$ electron operator $\psi_{el-\nu=1} = \exp(i\phi)$, we get RG-eq. for tunneling parameter $\Gamma_{\nu=1}$ in the same way. It is written as

$$\frac{d\Gamma_{ET-\nu=1}}{dl} = [1 - 2(\frac{1}{2})]\Gamma_{ET-\nu=1}. \qquad (15)$$

As the energy scale goes to small, $\Gamma_{ET-pf}$ decreases and $\Gamma_{QPT-pf}$ grows. $\Gamma_{ET-\nu=1}$ is invariant to the change in energy scale. By considering the RG flow, we can find the stable fixed point that corresponds to intermediate state. The configuration of two edges in this state is shown in Fig. 3. The Pfaffian edge behaves as the pinched-off channel and the $\nu = 1$ edge behaves as the connecting channel. Therefore, only the $\nu = 1$ edge contributes to the intermediate conductance, whose value is $e^2/h$. Next, we investigate the QPT in a non-perturbative regime.

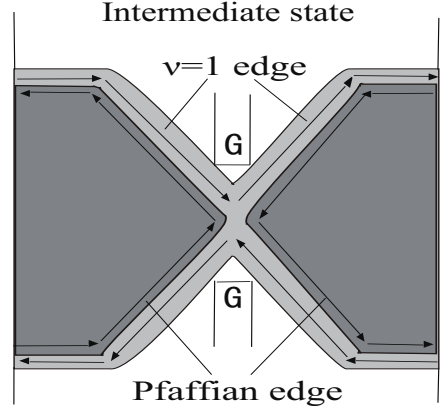

Fig. 3. (Color online) The intermediate edge state is represented. The Pfaffian edge is in the weak backscattering limit and the $\nu = 1$ edge is in the weak tunneling limit. Only the $\nu = 1$ edge channel contributes to the constant part of the two-terminal conductance.

Since the QPT term in eq. (9) is relevant, we should treat the quasi-particle tunneling non-perturbatively. To study this problem, we introduce an effective theory for the nonlinear degree of freedom $(\phi_\rho - \bar\phi_\rho) \equiv \theta$. The partition function of this system is written as

$$Z = \int D\phi_\rho D\psi_a$$
$$\exp(-S_B - S_F - \int d\tau H_{QPT})$$
$$= \int D\psi_a D\theta e^{-S_F} \int D\phi_\rho \delta(\theta - (\phi_\rho - \bar\phi_\rho))$$
$$\exp(-S_B - \int d\tau \Gamma_{QPT}\psi_{\frac{1}{2}} \cos(\frac{\phi_\rho - \bar\phi_\rho}{2}))$$
$$= \int D\psi_a D\theta e^{-S_F} \int D\phi_\rho \int D\lambda e^{(-i\int d\tau \lambda(\theta - (\phi_\rho - \bar\phi_\rho)))}$$
$$\exp(-S_B - \int d\tau \Gamma_{QPT}\psi_{\frac{1}{2}} \cos(\frac{\phi_\rho - \bar\phi_\rho}{2})),$$

where $S_F$ and $S_B$ are the actions for the free triplet Majorana fermion $\psi_a$ and free boson $\phi_\rho$, respectively. Integrating out $\phi_\rho$ and $\lambda$,[25] we obtain

$$Z = \int D\psi_a e^{-S_F} \int D\theta$$
$$\exp(-\sum_\omega \frac{|\omega|}{4\pi\nu}\theta(\omega)\theta(-\omega) - \int d\tau \Gamma_{QPT}\psi_{\frac{1}{2}} \cos(\frac{\theta}{2})).$$
$$(17)$$

This form is called the Caldeira-Legget model.[25–27] The first term is the friction term, and the second term can be regarded as a periodic potential for the $\theta$ field. The second term is interpreted as "magnetic field" in terms of the Ising model. In the non-perturbative regime, owing to the "magnetic field", which is proportional to $\cos\frac{\theta}{2}$, the saddle point solution of the spin field $\psi_{\frac{1}{2}}$ is no longer zero. $\theta$ has the value $4\pi n$ for

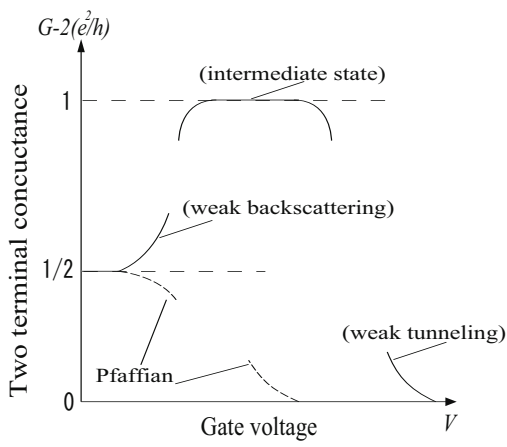

Fig. 4. (Color online) Schematic plot of the two-terminals conductance as a function of a negative gate voltage V. Solid curved lines denote the conductance of the anti-Pfaffian state and broken curved lines denote that of the Pfaffian state.

$\psi_{\frac{1}{2}} < 0$, while $\theta$ is $2\pi(2n + 1)$ for $\psi_{\frac{1}{2}} > 0$, where $n$ is an integer. The quasi-particle excitations in non-perturbative regime are described by instanton solutions similarly to the Laughlin state.[26, 27] In this situation, however, the form of the solution is different from the Laughlin state since the periodic potential is composed of two degrees of freedom. Ordinary instanton solutions are defined as a step from a potential minimum to another one. However in eq. (17), the reversal of the sign of the potential occurs in the tunneling process because of the operator $\psi_{\frac{1}{2}}$, hence the minima and maxima of the potential are exchanged. The instanton solution of such a process is called the half-instanton.[28] The tunneling dynamics of the Ising-class should be described by half-instanton solutions. By the instanton method, we construct an effective action for the non-perturbative regime. We introduce a function defined as

$$\frac{d\theta_{ins}}{d\tau} \equiv h(\tau), \quad (18)$$

where $\theta_{ins}$ is the solution for a single half-instanton. Thus the differential of the $n$ instanton solution and its Fourier transform are written as

$$\frac{d\theta_n}{d\tau} = \sum_{i=1}^{n} e_i h(\tau - \tau_i), \quad (19)$$

$$-i\omega\theta_n(\omega) = \sum_{i=1}^{n} e_i \tilde{h}(\omega) e^{i\omega\tau_i}, \quad (20)$$

where $e_i$ is the charge of the instantons, and $\tilde{h}(\omega)$ is the Fourier coefficient of $h(\tau)$. In particular,

$$\tilde{h}(\omega = 0) = \frac{1}{\sqrt{\beta}} \int d\tau h(\tau) = \frac{2\pi}{\sqrt{\beta}} \quad (21)$$

will be important. To derive the effective action, we consider a grand-canonical ensemble of the half-instantons. We neglect the $\omega$-dependence of $h(\omega)$ in the first term as

$$\sum_{\omega} \frac{|\omega|}{4\pi\nu} \theta(\omega)\theta(-\omega)$$

$$\approx \sum_{i,j}^{n} (\frac{\pi}{\nu\beta} \sum_{\omega} \frac{1}{|\omega|} e^{-i\omega(\tau_i - \tau_j)}). \quad (22)$$

The ground-canonical partition function is given as

$$Z = \int D\psi_a e^{-S_F} \sum_{n=0}^{\infty} \sum_{e_i}$$

$$\int_0^{\beta} d\tau_n \int_0^{\tau_{n-1}} \cdots \cdots \int_0^{\tau_1} d\tau_0 \prod_{i=0}^{n} z(\tau_i)$$

$$\exp[-\sum_{i,j}^{n} (\frac{\pi}{\nu\beta} \sum_{\omega} \frac{1}{|\omega|} e^{-i\omega(\tau_i - \tau_j)}) e_i e_j], \quad (23)$$

where $z(\tau) = \omega_0 e^{-S_{ins}}$ is the fugacity of the instanton and $\omega_0$ is the characteristic energy of the tunneling process. The value must be directly proportional to the energy operator $\epsilon_a = \psi_a \bar{\psi}_a$, i.e. $\omega_0 = z_0 \epsilon_a(\tau)$, where $z_0$ is a constant. Introducing the Stratonovich-Hubberd field $\Theta$, we obtain

$$Z = \int D\psi_a e^{-S_F} \sum_{n=0}^{\infty} \sum_{e_i} \frac{1}{n!}$$

$$\int_0^{\beta} d\tau_n \int_0^{\beta} \cdots \cdots \int_0^{\beta} \prod_{i=0}^{n} \epsilon(\tau_i)$$

$$\int D\Theta \exp(-\sum_{\omega} \frac{|\omega|\nu}{4\pi} \Theta(-\omega)\Theta(\omega) + \frac{i}{\nu} \sum_{i} e_i \Theta(\tau_i))$$

$$= \int D\psi_a e^{-S_F} \int D\Theta \exp(-\sum_{\omega} \frac{|\omega|\nu}{4\pi} \Theta(-\omega)\Theta(\omega)$$

$$-2z_0 \int d\tau \psi_a \bar{\psi}_a \cos(\Theta(\tau)/\nu)). \quad (24)$$

Note that the second term is the electron tunneling term. As the cases of the Laughlin and Pfaffian states, we can treat the quasi-particle tunneling as the weak electron tunneling effectively in the strong coupling regime. In other words, the quasi-particle and electron tunneling processes are dual with each other. Therefore, we can easily calculate the temperature dependence of two-terminals conductance,[11–13, 29] which is

$$G \propto T^4. \quad (25)$$

In this paper, we have studied the anti-Pfaffian edge tunneling in the single point-contact system. We have expected the intermediate edge state and calculated the conductance. We have also shown the duality between the weak coupling electron and strong coupling quasi-particle tunneling processes by the instanton method. The schematic portrait of the two-terminal conductance is shown in Fig. 4 as a function of gate voltage. The solid curves show the conductance of the anti-Pfaffian state and the broken curves show the conductance of the Pfaffian state. This figure clearly shows the difference in gate voltage dependence between the two states. From this result, one can distinguish the anti-Pfaffian state from the Pfaffian state experimentally in a single point-contact system.